\documentclass[prd,twocolumn,epsfig]{revtex4}
\usepackage{graphicx}
\newcommand{\be}{\begin{equation}}
\newcommand{\ee}{\end{equation}}

\newcommand{ \K }{ \raisebox{-0.8ex} {\scriptsize \it K} }

\begin{document}

\title{Simple condensation of composite bosons  in a number conserving approach
to many fermion systems }


\author{Fabrizio Palumbo}
\affiliation{%
  {\small\it INFN -- Laboratori Nazionali di Frascati}  \\
  {\small\it P.~O.~Box 13, I-00044 Frascati, ITALIA}          \\
  {\small e-mail: {\tt fabrizio.palumbo@lnf.infn.it}}     
}%




\begin{abstract}

We recently derived the Hamiltonian of fermionic composites by an exact  procedure of  bosonization. In the present paper  expand this Hamiltonian in the inverse of the number of fermionic states in the composite wave function and give
the necessary and sufficient conditions for the validity of such an expansion. We compare the results to the Random phase Approximation  and 
the BCS theory and perform an illustrative application of the method.


\end{abstract}

\maketitle

\clearpage


The low energy physics of several many fermion systems is determined by the formation of composites, accompanied or not by their
condensation.   Examples of simple condensation (ground state  built by a large number of composite bosons in one single-particle state while all other occupation numbers are of order 1) can be found in  superconducting materials,
ultracold fermionic atoms in magnetic traps~\cite{Holl},  ultrasmall metallic grains~\cite{Ralp} and atomic nuclei~\cite{Bohr,Arim}. Examples of
multiple condensation (ground state  built by a large number of bosons in 2 or more single-particle states) essentially occur in spatially separated systems as in the Josephson effect and superleak or in systems whose components differ by some internal quantum number of nongeometrical nature~\cite{Legg}. The theory of all such systems requires some bosonization procedure.  For finite systems  it is important to respect   particle number conservation,
and this becomes unavoidable in the study of  effects related to the  total number of fermions, its parity and supersymmetry~\cite{Iach}. But respecting this symmetry can be important also for infinite systems~\cite{Legg}. 

There is an infinite literature on this subject which we cannot review here, but to our kwnoledge there is no unified mehtod which 
 respects fermion number conservation and treats ground and excited states on the same footing.
 We recently proposed an  exact bosonization procedure which, as such, implements these features.  Starting from the operator form of the partition function,  after discretization of the Euclidean time 
we perform an independent Bogoliubov transformation at each time slice. The time
 dependent parameters of the transformation are then associated with the fields of dynamical
 composite bosons  in the presence of quasiparticles satisfying  a compositeness condition which avoids double-counting~\cite{Palu}. 
This formalism  has been applied also to relativistic field theories, including gauge theories, at zero
and finite temperature and fermion density~\cite{Cara}. 

 In its  first formulation~\cite{Palu1} the method was restricted to complete  bosonization
 (absence of unpaired fermions). The resulting Hamiltonian was studied  by an expansion in inverse powers of {\it $\Omega$, the index of nilpotency of the composites, which is the number of fermionic states in their structure functions}, assuming  simple
 condensation with one-dimensional condensation sector. The condensation sector is  the Fock subspace of composite bosons which
contains the condensed boson and all other bosons whose mixing with the condensed one is allowed by symmetries. We call free space the complementary subspace.  
 In the present work we  generalize the results of Ref. ~\cite{Palu1} to a condensation sector of arbitrary dimension.  
 In this framework we can study  BCS-BEC transitions in ultracold systems in magnetic traps~\cite{Grei},  effects related to the number of fermions and the parity of this number,  shape transitions and
 shape coexistence~\cite{Wood} in atomic nuclei. One of the motivations of our work is indeed to derive  the phenomenological Interacting Boson Model~\cite{Arim} of atomic nuclei  starting from an effective nuclear Hamiltonian in a model space.  
  Fermion Hamiltonians in a model space can 
be written in the form 
\be
H_{F} = {\hat  c}^{\dagger} ( e + {\mathcal M}) \, {\hat c} -
{ 1\over 4}  g\K    \, {\hat c}^{\dagger} \, F_K^{\dagger} \,{\hat  c}^{\dagger} \,\,  {\hat c} \, F_K \,{\hat  c}\,.
\ee
 $\hat{c}^{\dagger}, \hat{c}$ are fermion creation-annihilation operators and 
we adopt a summation convention over repeated
 indices and a matrix notation, for instance 
 ${\hat c} \, F_K \,{\hat  c} = \sum_{m_1m_2} {\hat c}_{m_1}\, (F_K)_{m_1m_2} \,{\hat  c}_{m_2}$.
 $ e $ is the matrix of single-fermion energy, ${\cal M}$ the matrix of any interaction with external fields and  $F_K $  are the potential form factors normalized according to
$ \mbox{tr} ( F_{K_1}^{\dagger} F_{K_2}) = 2 \, \Omega \delta_{K_1 K_2}$. The separable terms of the potential are assumed of particle-particle type: the particle-hole ones must be  rearranged in this form. The  Hamiltonian of composite bosons ~\cite{Palu1} is
  \begin{eqnarray}
& &H_B= : \mbox{tr} ( R  \,{\mathcal B}^{\dagger}   e\, {\mathcal B})    - 
{ 1 \over 4} \sum_K g_K  \mbox{tr} (R \, {\mathcal B}^{\dagger}  F_K^{\dagger}) \,
\mbox{tr}(R F_K {\mathcal B}) 
\nonumber\\
&& +{ 1\over 2} \sum_K g_K  \mbox{tr} \left[(R -1) F_K^{\dagger} \, F_K
- R\, {\mathcal B}^{\dagger}  F_K^\dagger R F_K {\mathcal B} \right] :  \label{Hinitial}
\end{eqnarray}
where the colons mean normal ordering,  $
{\mathcal B} ={1 \over \sqrt{\Omega}} \, \, {\hat b}_J B_J^{\dagger} 
\,,  \,\,\,R= \left( 1\!\!1 + {\mathcal B}^{\dagger} {\mathcal B}\right)^{-1}  $
and we omitted the interaction with external fields $H_I=  :\mbox{tr} (R \,{\mathcal B}^{\dagger}   {\mathcal M}\, {\mathcal B}):$ .
In the above equations the  ${\hat b}_K$ are destruction operators of composites with quantum numbers $K$ and
the matrices $B_K$ are their structure functions. Unlike the composite  operators
$ {\hat B}_K= { 1 \over 2 {\sqrt \Omega }} \,{\hat c} \, B_K \,{\hat  c}$, they obey canonical commutation relations
provided the conditions on the structure functions reported below~(\ref{normalization1},\ref{normalization2}) are satisfied.
 In a sector of $n_F$ fermions $H_B$ must used with the constraint~\cite{Palu1}
\be
\mbox{tr} \left( R  \,{\mathcal B}^{\dagger} \, {\mathcal B} \right) = n_F \,. \label{fermion-number}
\ee
  We denote the quantum numbers of the composite bosons in the
  condensation/free  sector by  $c_i$ /$f_i$ respectively. 
 {\it We  call "zero" the condensation mode and  $\sigma$ the remaining states in the condensation sector}.  We will use the index ${\overline K} $ to represent  both $\sigma$- and $f$-bosons. 
 
 The difficulty dealing with the bosonic Hamiltonian comes from the operator $R$ which is a nonpolynomial
 function  of bosonic creation-destruction operators. If the number of composite bosons is much smaller
 than $\Omega$ we can overcome the difficulty by expanding $R$ with respect to $ {\mathcal B}^{\dagger} {\mathcal B}$.
 Otherwise a subtraction of the condensate is necessary before an expansion can be done~\cite{Palu1}.
  Because only the "zero"mode will have 
 a large occupation, 
we can make  the  number conserving approximation
$ \Omega^{-1} \,
\hat{b}_0^{\dagger}\hat{b}_0  \approx  n / \Omega = \nu_0, \,\,
 $ where $n= { 1\over 2}n_F$ is the number of bosons.  
Now we can write  the normalization conditions
\be   
{ 1 \over 2 \Omega} \mbox{tr} \, C_{00}  =1,
{ 1 \over 2 \Omega} \mbox{tr} \left[ C_{{\overline K}_1 {\overline K}_1} -  \nu
 C_{{\overline K}_100 {\overline K}_1}\right] =
 \delta_{{\overline K}_1 {\overline K}_1} \label{normalization1}
 \ee
 \be 
 { 1 \over 2 \Omega} \mbox{tr} \, C_{0\sigma}=
{ 1 \over 2 \Omega} \mbox{tr} \, C_{\sigma0}  =0                                                                                                                                                                         
\label{normalization2}
  \ee
where $
 C_{K_1 K_2K_3K_4...}= \Gamma B_{K_1} B_{K_2}^{\dagger} \Gamma B_{K_3} B_{K_4}^{\dagger} ...$ and $
\Gamma= \left( 1+\nu B_0B_0^{\dagger}\right)^{-1} $.
They are derived as in  Ref.\cite{Palu1} and coincide with those reported there if the condensation sector is one-dimensional. 
The subtraction parameter $\nu$ (a priori different from $\nu_0$) is  determined by the constraint~(\ref{fermion-number})
which for simple condensation using Eqs.~(\ref{normalization1},\ref{normalization2}) becomes
$ n\, (\nu_0 - \nu) = \mbox{terms of order} \,\,\Omega^0 $.
This equation is satisfied by $\nu = \nu_0 +\mbox{terms of order} \,\,\Omega^{-1} $ which give  contributions of order 1 which depend on the number of fermions but not on their state. 
Following Ref.\cite{Palu1} $H_B$
can be expanded in powers of $ \Omega^{-{1\over 2}}$.
 Neglecting terms of order  $ \Omega^{-{1\over 2}}$ it becomes
   \begin{eqnarray*}
   H_B &=& {\mathcal E}_{00} \, n +  {\mathcal E}_{0\sigma} \,b_0^{\dagger} b_{\sigma} + 
    {\mathcal E}_{\sigma 0}\,b_{\sigma}^{\dagger} b_0 +
    {\mathcal E}_{{\overline K}_1 {\overline  K}_2}\,  b_{{\overline K}_1}^\dagger b_{{\overline K}_2}
   \nonumber\\
    &+&  {\mathcal E}_0+  { 1\over 2} \left(
     {\mathcal W}_{{\overline K}_1 {\overline  K}_2}  { 1 \over \Omega}\,b_0 b_0 \,b_{{\overline K}_1}^\dagger 
    b_{{\overline K}_2}^\dagger  + H.c. \right) 
    \end{eqnarray*}
where
\begin{eqnarray}
{\mathcal E}_0&=& \nu_0T_{0000}- \Omega g_c \nu_0 (D_{000c}   D_{0c}^*+cc)
\nonumber\\
&-& \Omega g_K \nu_0 { 1\over 2 \Omega}{ \mbox tr} \left[\Gamma B_0(B_0^\dagger F_K^\dagger F_K+F_K^\dagger \Gamma F_K B_0^\dagger)\right]
\nonumber\\
{\mathcal E}_{00}&=& 2 \,{\overline e} +T_{00} -  \Omega g_c | D_{0 c}|^2
 \nonumber\\
 {\mathcal E}_{0\sigma} &=& {\mathcal E}_{\sigma0} ^*= T_{0\sigma} - \nu_0 \,T_{0\sigma00}  - \Omega g_c 
 \left[D_{0c} D_{\sigma c}^* \right.
\nonumber\\
&+ &    \left.  - \nu_0 ( D_{0  c}^*D_{0\sigma 0 c}  + D_{0  c}D_{\sigma0 0 c}^* )\right]
 \nonumber\\
 {\mathcal E}_{{\overline K}_1 {\overline K}_2} &=&{\mathcal E}_{{\overline K}_2 {\overline K}_1}^* =  \nu_0 \left[T_{0000} - g_c \Omega ( D_{000c}D_{0c}+cc)\right] \delta_{{\overline K}_1 {\overline k}_2}
 \nonumber\\
 & +& T_{{\overline K}_1 {\overline K}_2} -  \nu_0
\left[ T_{{\overline K}_1{\overline K}_200} + \,T_{{\overline K}_1 00{\overline K}_2}  +T_{0{\overline K}_2 {\overline K}_1 0}
\right.
\nonumber\\
&-& \left.  \nu_0 \left( T_{0 {\overline K}_2 {\overline K}_1000}+T_{{\overline K}_100 {\overline K}_200} \right) \right]
\nonumber\\
&+&  \Omega g_c  \,  \nu_0 \, \left\{   \left[ D_{{\overline K}_1 {\overline K}_20c} \, D_{0c}^*  + 
\, D_{0c} \, D_{0{\overline K}_1 {\overline K}_2c}^* 
\right. \right. 
\nonumber\\
&- & \left. \left. \nu_0 ( D_{0{\overline K}_2 {\overline K}_100c} \, D_{0c}^* +
 D_{{\overline K}_100 {\overline K}_20c}  D_{0c}^*)   \, \right]  +cc  \right\}
\nonumber\\
&-& \Omega g_K   \, \nu_0 \left[   D_{{\overline K}_1K  } D_{{\overline K}_2K }^* 
- \nu_0 (  D_{{\overline K}_100 K  } D_{{\overline K}_2K }^*          + cc  ) 
\right.
 \nonumber\\
 &+ & \left. \nu_0^2 (  D_{{\overline K}_100K  } D_{{\overline K}_200K }^* + D_{0{\overline K}_20K  } D_{0{\overline K}_10K }^*    ) \right]
 \nonumber\\
{\mathcal W}_{{\overline K}_1 {\overline  K}_2} &=&- 2 \, \left\{ T_{{\overline K}_10{\overline K}_20} -  \nu_0 \, T_{{\overline K}_10{\overline K}_2000} 
\right.
\nonumber\\
& +& 
\left.  \Omega g_c \left[  - D_{0c}^* D_{{\overline K}_10{\overline K}_2c}  + \nu_0 \left(  D_{0c}^* D_{{\overline K}_10{\overline K}_200c} 
\right. \right. \right.
\nonumber\\
&+&\left. \left. \left. D_{0{\overline K}_10{\overline K}_20c}^* D_{0c} \right) \right] 
+  \Omega g_K \left(  - D_{0{\overline K}_10K}^* D _{{\overline K}_2 K}  \right. \right. 
\nonumber\\
&+& \left.   \left.\nu_0 D_{0{\overline K}_10K}^* D_{{\overline K}_200K} \right)  \right\}\,.
 \end{eqnarray}
We used the definitions
\begin{eqnarray}
  T_{K_{1}K_{2}...K_{2l-1}K_{2l}}& =& { 1\over  \Omega} \mbox{tr}\left[  C_{K_{1}K_{2}}... 
  C_{K_{2l-3}K_{2l-2} }\right. 
  \nonumber\\
  && \left.
    \times \Gamma B_{K_{2l-1}} \,( e - {\overline e})\, B^{\dagger}_{K_{2l}}  \right]
  \nonumber\\
   D_{K_{1}K_{2}...K_{2l-1}K_{2l}}& =& { 1\over 2 \Omega} \mbox{tr}\left[  C_{K_{1}K_{2}} ...   C_{K_{2l-3}K_{2l-2} }\right. 
   \nonumber\\
   & & \left.   
  \times  \Gamma B_{K_{2l-1}} F_{K_{2l}}^{\dagger} \right]
    \end{eqnarray}
    where $l$ is any integer. The above equations are valid for an arbitrary choice of ${\overline e}$ as a consequence of
    the constraint~(\ref{fermion-number}) on the number of fermions. All the constants  appearing in  $H_B$ are of order 
    $\Omega^0$. Then assuming $n \sim \Omega$ the first term is of order $\Omega$, the  second and third  are of order $\Omega^{{ 1\over 2}}$
   (because  the matrix elements of the 
  operators $ {\hat b}_0^{\dagger} \,{\hat b}_{\sigma}$ are of order $\sqrt n$), and the other terms are of order $ \Omega^0$. Estimating the order of the different
    terms we assume that the coupling constants scale according to $g_K \sim \Omega^{-1} \delta e$, where $  \delta e$
    is the single fermion energy spreading. This assumption is necessary  to get finite energies in infinite systems.            
   Two comments are in order:  i)
     all the constants appearing in $H_B$ vanish unless both quantum numbers ${\overline K}_1{\overline K}_2$ belong  either to the condensed  or to the free sector because of different symmetries in these sectors, ii)
  the zero-boson has a strong mixing with  the $\sigma$-bosons due to the terms of order $\Omega^{{1\over 2}}$. But if it is the true
  condensed boson  such mixing must be absent
\be 
{\mathcal E}_{0\sigma}={\mathcal E}_{\sigma0}=0\,.   \label{decoupling}
\ee
 As we will see this decoupling prevents the occurrence 
of boson-particle-boson-hole states in the ground state, and it is therefore similar to an ordinary selfconsistency condition.
   
We can now
derive a number of results without numerical calculations by introducing phonon operators. This can be done in two steps. First  we introduce the phonon operators $
{\hat A}_{{\overline K}}={ 1\over  \sqrt{n}} \, \hat{ b}_{{\overline  K}} \, \hat{b}_0^{\dagger}$
and rewrite the Hamiltonian accordingly
\begin{eqnarray}
\hat{H}_B  & = & \mathcal{E}_0 +
 \mathcal{E}_{0 0}  \, n+
{\mathcal E}_{{\overline K}_1{\overline K}_2} \hat{A}_{{\overline K}_1}^{\dagger}\hat{A}_{{\overline K}_2} 
\nonumber\\
&+& \nu_0 \left( \,  {\mathcal W}_{{\overline K}_1 {\overline  K}_2} {\hat A}_{{\overline K}_1}^\dagger {\hat A}_{{\overline K}_2}^\dagger   + H.c. \right) \,.
\end{eqnarray}
This is the Bogoliubov Hamiltonian for a superfluid bosonic system and can be diagonalized by a bosonic Bogoliubov transformation $
  {\hat A}^{\dagger}=  {\hat \Phi}^{\dagger} U^\dagger+  {\hat \Phi} V^\dagger \,, \,\,\,
  {\hat A}= U {\hat \Phi}+ V{\hat \Phi}^{\dagger}$.
  This  transformation becomes simple when, due to symmetries,  there is a one-to-one correspondence 
between the quantum numbers  ${\overline K}_2$ and $ {\overline K}_1$:  ${\overline K}_2=  {\overline K}_2 ({\overline K}_1)$. This is guaranteed  only  for the free sector, but we
assume it to be true for all noncondensed modes in order to illustrate some general features.  Then  $
{\mathcal E}_{{\overline K}_1{\overline K}_2}={\mathcal E}_{{\overline K}_1} \delta_{{\overline K}_1{\overline K}_2}\,, \,\,\,
 {\mathcal W}_{{\overline K}_1 {\overline K}_2}= {\mathcal W}_{{\overline K}_1} \delta_{
  {\overline K}_2 ({\overline K}_1), {\overline K}_1}$.
  After transformation $H_B$ takes its final form
 \be
H_B =\mathcal{E}_0+\sum_{{\overline K}} \sqrt{ E_{{\overline K}}  - {\mathcal E}_{{\overline K}}} +   \mathcal{E}_{0 0}  \, n
 + E_{{\overline K}} \,\, {\hat \Phi}_{{\overline K}}^{\dagger}\, \,{\hat \Phi}_{{\overline K}}   \label{phonons-2} \label{Hfinal}
\ee
where $  E_{{\overline K}} = \sqrt{ {\cal E}_{{\overline K}}^{2} - \nu_0^2  \, {\mathcal W}_{{\overline K}} ^{2}}$.  The ground state   is the vacuum of the phononic operators ${\hat \Phi}$, which written in terms of the bosonic composites is
\begin{eqnarray*}
| GS \rangle = | \exp \left(  { 1\over 2n}   \,   b_{{\overline K}_1}^{\dagger}\left( V  (U^*)^{-1}\right)_{{\overline K}_1{\overline K}_2}   
b_{{\overline K}_2}^{\dagger} {\hat b}_0 \, {\hat b}_0 \right)
\left(\hat{b}_0^{\dagger}\right)^n |0 \rangle \,.
 \end{eqnarray*}
 $|GS\rangle $
is a coherent state of  phonon pairs which are 2p-2h boson states and therefore 4p-4h fermion states built on the $b_0$ condensate, at variance~\cite{Ring}  with the RPA. 
The above results have been derived under the condition that the occupation of noncondensed bosons be of order 1
  \be
< GS|  \sum_{{\overline K}}{\hat b}_{{\overline K}}^{\dagger} \,{\hat b}_{{\overline K}} |GS> = 
{ 1 \over 2}\sum_{{\overline K}}\left( { {\cal E}_{{\overline K}} \over  E_{{\overline K}}} - 1 \right)\  << \Omega
\,.  \label{average-number}
\ee
 {\it This is the necessary and sufficient condition for the validity of the assumption of simple condensation and
 of the $\Omega^{-1}$ expansion}. So this is also the condition for condensation in a finite system.
 
 The structure functions $B_K$ must be determined by a variational calculation on ground state and phonon energies under the 
 normalization and decoupling conditions. This can be done in 
 the following way.
We write $B_0 = {\overline B}_0 + { 1\over \Omega} \delta B_0$. We solve the variational equation for $ {\overline B}_0 $ neglecting the first 2 terms of (\ref{Hfinal}) which are of order 1 
and use the result to determine the correction $\delta B_0 $. In the present approximation in which we neglect terms of order $\Omega^{-{1\over 2}}$ this correction will alter only the contribution of the term $ {\mathcal E}_{00}n$. A comparison with the BCS theory can clarify the
meaning of this procedure.
 The parameter $  {\mathcal E}_{00}$ is indeed the BCS energy per particle in the form of the quasichemical equilibrium theory~\cite{Blat}. To see this we must sketch an essential point in the derivation of $H_B$. 
  The  time-dependent Bogoliubov transformation changes  the fermion vacuum into the quasiparticle vacuum 
$
| {\mathcal B} \rangle = \exp \left( b_K {\hat B}_K^{\dagger}\right) |0 \rangle 
$,
where the  ${\hat B}_K = { 1 \over 2 {\sqrt \Omega }} \,{\hat c} \, B_K \,{\hat  c}$ are composite operators and the  $b_K$
time-dependent holomorphic variables. The partition function involves an integral  over these   variables  which acquire 
 the meaning of dynamical bosonic  fields with fermion number 2, so that all the terms in the expansion of
$| {\mathcal B} \rangle  $ have fermion number zero. A further transformation of the partition function into a trace over
a bosonic Fock space leads to the expression~(\ref{Hinitial}). The coherent states
  $|{ \mathcal B} \rangle$  can be identified with the BCS states setting 
  \begin{eqnarray*}
 { v_m \over u_m} =   { b_0 \over {\sqrt \Omega}} \,\,  (B_0)_{m,-m} \label{BCS}\,, \,\,\, \,\,
   u_m= \left[ 1 +{ |b_0|^2 \over \Omega}(B_0)_{m,-m}^2\right]^{-{1\over 2}}
     \end{eqnarray*}
  with standard BCS notation.
   But if one is ready to break fermion number conservation one can set $b_0^2 \approx n$. Then 
 the BCS normalization $u_m^2+v_m^2=1$ is identically enforced and the first of the normalization conditions~(\ref{normalization1}) becomes the BCS condition on fermion number
$ \sum_m  v_m^2 ( 1 + v_m^2)^{-1} = n_F$.

  One can easily check that  the variation of ${\mathcal E}_{00}$ with respect to $B_0$ yields the standard gap equation. But this equation  must now be solved under the decoupling conditions~(\ref{decoupling}).These conditions are obviously absent  when the condensation sector is onedimensional, in which case  one can use results obtained in the BCS theory to evaluate the phonon energies and $\delta B_0$ which  corresponds to the contribution arising from projection of the BCS wave function (but it includes the contribution of  states in the free sector). 
   \begin{figure}[htbp]
     \includegraphics[height=5cm]{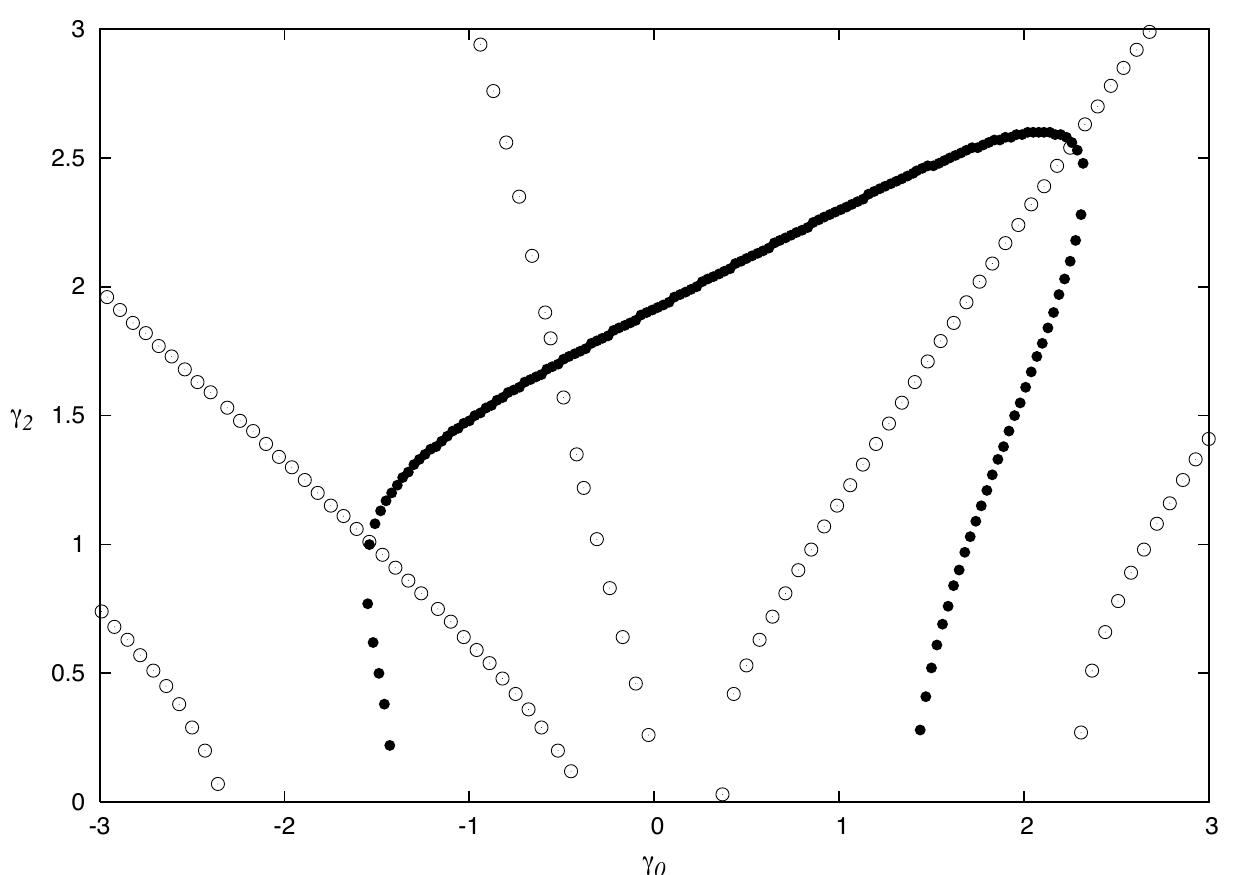}
     \caption{Contour plots of the  first of normalization conditions (\ref{normalization1})(full circles)  and of the decoupling condition 
     (\ref{decoupling}) (empty circles) for $ \xi=1.5 $ and $ \nu_0 = 0.5$. The minimum energy is for a spherical shape, at $\gamma_2
     =0,   \gamma_0=1.4$.}
     \label{}
     \hspace{-0.5cm}\includegraphics[height=5cm,width=7.5cm]{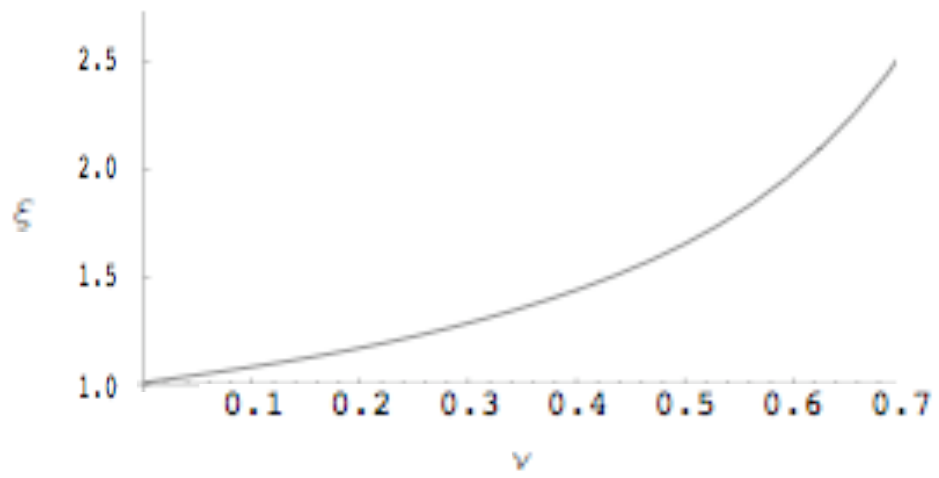}
     \caption{The critical line separating the deformed phase (upper part) from the spherical one (lower part).}
     \label{}
\end{figure}
We now show how  our method can be applied in practice to the study of shape transitions in nuclei and the 
 BEC-BCS transition at zero temperature, which can be regarded as the transition from molecular to Cooper structure
 of the condensed boson. In fact the convenience of the quasichemical equilibrium theory in this context has been advocated long ago~\cite{Legg1}.
  In the principal frame of inertia of the nucleus the zero-mode will be a superposition of  $b_{LM}$-bosons with $M=0$, the $\sigma$-modes will be  all the other
independent superpositions of modes with $M=0$ and the  free modes will have $M\neq 0$. To determine the structure functions
we must minimize the ground state  and the phonon energies under  the decoupling  and normalization conditions~(\ref{decoupling},\ref{normalization1}, \ref{normalization2}). The problem simplifies considerably if we assume the condensation sector to be 2-dimensional, including only the states  $L=0,2$, the $s$- and $d$- bosons of the Interacting boson model. As a further simplification  we assume  the nucleons  to live in a single $j$-shell and have  constant single-particle energy and monopole and quadrupole 
   paring  interactions. The energy is therefore a function of the ratio of couplings $\xi= g_2 / g_0$.
 In this model space $\Omega =j+{ 1\over 2}$ and the potential form factors are proportional to Clebsh-Gordan coefficients
$
 \left(F_{LM} \right)_{m_1m_2}= <jm_1jm_2|LM> \sqrt{2j+1}\,.
 $
  Then the  structure matrices  in the condensation sector can be parametrized according to $
B_0 = \gamma_0 F_{00} + \gamma_2 F_{20}\,,  \,\,\, B_{\sigma}= {\mathcal N} ( \cos \beta F_{00} + \sin \beta F_{20}) $,
and all the parameters are determined by the constraints. 
 $ \beta$ and ${\mathcal N} $, are determined by  Eqs.~\ref{normalization1}. Eqs. (\ref{normalization2}) and (\ref{decoupling}) 
 give rise respectively to a closed and open contour lines in the $\gamma_0, \gamma_2$ plane as shown in Fig.1 where only the upper part is shown (the lower one can be obtained 
 by a parity inversion). The intersection for which ${\mathcal E}_{00}$ is minimum determines the values of the
 $\gamma$-parameters and therefore the shape of the nucleus.  In Fig. 2 we plot the critical line which separates the deformed from the spherical shape. When $\nu_0$ approaches 1 the shell is fully occupied, and because of the Pauli principle
 the bosonic approximation cannot be valid. 
However we have a tentative interpretation of the steep rise of the critical line in this region. For $\nu_0=1$ there is a
 unique nuclear state which is spherical. Therefore approaching 1,  larger and larger values of the strength of the quadrupole pairing are necessary
 to get a deformed shape.
 Below the critical line the ground state is a condensate of bosons with zero
angular momentum  ({\it seniority limit}).  The  condensation sector is one-dimensional and 
\be
   \hat{H}_B = 2n \,e +  g_0 n^2 - \Omega g_0 n + E_{L} \hat{A}_{LM}^{\dagger}  \hat{A}_{LM} 
 \ee
 where 
\be
E_{L} = \Omega g_0 | 1-\xi |  \sqrt{ 1+ {1 \over1-\xi}  \left[ 1 - 4 \left( { 1 \over 2} - \nu \right)^2   \right]  } \,.
\ee
  Note that the Hamiltonian is expressed in terms of the $A$-phonons, because the coefficients ${\mathcal W}$ vanish.
The phonon  energies   are even functions of the deviation from half filling and vanish for  $\xi=1 $, close to the critical line.
For $\xi=0$ we reproduce the spectrum of the pairing model~\cite{Ring}. 
Inclusion of a nonconstant single-particle energy does not require any modification of the procedure outlined. Instead 
if we increase  the model space and/or the condensation sector,  the structure functions will depend on more parameters and we will have to solve 
a constrained gap equation. The resulting Hamiltonian can be directly compared to that of the Interacting Boson model in the intrinsic frame.
 
\vspace{-0.5cm} \subsection *{Acknowledgment}

We are grateful to K. Hayakawa and A. Sportiello for interesting discussions of numerical methods.


\begin{thebibliography}{11}

\bibitem{Holl}
J. Holland, B. de Marco and D. S. Jin, Science, 285 ( 1999) 1703

\bibitem{Ralp}
D. C. Ralph, C. T. Black and M. Tinkham, Phys. Rev. Lett. 76 ( 1996) 1767; 78 (1997) 4087

\bibitem{Bohr}
A. Bohr, B. R. Mottelson and D. Pines, Phys. Rev. 110 ( 1958) 936

\bibitem{Arim}
F. Iachello and A. Arima, The Interacting Boson Model, Cambridge University Press, Cambridge, 1987


\bibitem{Legg}
A. J. Leggett, Rev. Mod. Phys. 73 (2001)307

\bibitem{Iach}
F. Iachello and P. Van Isacker, The interacting boson-fermion model, Cambridge University Press, Cambridge 1991

\bibitem{Palu}
F. Palumbo, [Proceedings of 9th International Spring Seminar on Nuclear Physics: Changing Facets of Nuclear Structure, Vico Equense, Italy, 20-24 Ma 2007; [arXiv: nucl-th/0711.4911].

\bibitem{Cara}
S. Caracciolo, V. Laliena and F. Palumbo,   JHEP {\bf 0702} (2007) 034
  [arXiv:hep-lat/0611012].
F. Palumbo, Phys. Rev. D 78 (2008) 014514; S. Caracciolo, F. Palumbo
and G. Viola, Annals Phys. doi:10.1016/j.aop.2008.10.006,  arXiv:0808.1110 [hep-lat].

\bibitem{Palu1}
F. Palumbo, Phys. Rev. C72 (2005)014303;
Eur. Phys. J. B 56 (2007) 335 

\bibitem{Grei}
M. Greiner, C. A. Regal and D. S. Jin, Nature (London) 426 ( 2003) 537; S. Jochim et al., Science 302 ( 2003) 2101;
M. W. Zwierlein et al., Phys. Rev. Lett. 91 (2003) 250401

\bibitem{Wood}
J. L. Wood et al.. Phys. Rep. 215 (1992) 101

\bibitem{Ring}
P. Ring and P. Schuck, The Nuclear Many Body Problem, Springer-Verlag, Berlin, 1980


\bibitem{Blat}
J. M. Blatt, Theory of superconductivity, Academic Press, New York and London (1964)

\bibitem{Legg1}
A. J. Leggett, in Modern Trends in the Theory of Condensed Matter, edited by A. Pekalski and R. Przystawa (Springer-Verlag, Berlin, 1980)




\end{thebibliography}
\end{document}